\documentclass[preprint,aps,showpacs,preprintnumbers,amsmath,amssymb]{revtex4}  

\usepackage{graphicx}
\usepackage{dcolumn}
\usepackage{bm}

\begin{document}

\title{Pulse-Bandwidth Dependence of Coherent Phase Control of Resonance-Mediated (2+1) Three-Photon Absorption}

\author{Andrey Gandman, Lev Chuntonov, Leonid Rybak, and Zohar Amitay}
\email{amitayz@tx.technion.ac.il} %
\affiliation{Schulich Faculty of Chemistry, Technion - Israel
Institute of Technology, Haifa 32000, Israel}


\begin{abstract}
We study in detail coherent phase control of femtosecond
resonance-mediated (2+1) three-photon absorption and its dependence
on the spectral bandwidth of the excitation pulse. The regime is the
weak-field regime of third perturbative order. The corresponding
interference mechanism involves a group of three-photon excitation
pathways that are on resonance with the intermediate state and a
group of three-photon excitation pathways that are near resonant
with it. The model system of the study is atomic sodium (Na), for
which experimental and numerical-theoretical results are obtained.
Prominent among the results is our finding that with simple proper
pulse shaping an increase in the excitation bandwidth leads to a
corresponding increase in the enhancement of the three-photon
absorption over the absorption induced by the (unshaped)
transform-limited pulse. For example, here, a 40-nm bandwidth leads
to an order-of-magnitude enhancement over the transform-limited
absorption.
\end{abstract}

\maketitle

\section{\label{sec:introduction}Introduction}

Coherent control utilizes the coherent nature of the light in order
to control transition probabilities of the irradiated quantum system
to desired target states
\cite{tannor_kosloff_rice_coh_cont,shapiro_brumer_coh_cont_book,warren_rabitz_coh_cont,rabitz_vivie_motzkus_kompa_coh_cont,dantus_exp_review1_2}.
Due to their broad spectrum, femtosecond pulses provide a wide
coherent band of such photo-induced pathways. The control is
achieved by manipulating interferences among multiple
initial-to-final state-to-state pathways. Constructive interferences
enhance the transition probability, while destructive interferences
attenuate it. The interferences are manipulated by shaping the
pulse, i.e., manipulating the phase, amplitude, and/or polarization
of its various spectral components. 

Prominent among the processes, over which such femtosecond control
has been demonstrated to be very effective, are multiphoton
processes that are of fundamental as well as applicative importance
for nonlinear spectroscopy and microscopy
\cite{dantus_exp_review1_2,silberberg_2ph_nonres1_2,
baumert_2ph_nonres, silberberg_2ph_1plus1,
girard_2ph_1plus1,becker_2ph_1plus1_theo,
dantus_2ph_nonres_molec1_2,leone_res_nonres_raman_control,
silberberg_antiStokes_Raman_spect,leone_cars,amitay_3ph_2plus1,
gersh_murnane_kapteyn_Raman_spect,taylor_raman_imaging,
amitay_2ph_inter_field1,silberberg-2ph-strong-field,
weinacht-2ph-strong-theo-exp,wollenhaupt-baumert1_2}.
For a given energy, manipulating the pulse shape manipulates the
cross-section of the corresponding process.
For rational shaping of femtosecond pulses, which are favorable for
a desired result, the most convenient excitation regime is the
weak-field regime where the photo-excitation is validly described
within the time-dependent perturbation theory of the lowest non-vanishing order.
For N-photon absorption processes it is the N$^{th}$ order.
Such a description allows a transformation of the photo-excitation
picture into the frequency domain, where the interfering pathways
and their interference mechanism can be identified and serve as a basis
for the rational pulse shaping.
Corresponding examples of past control studies include the control
over non-resonant two-photon absorption
\cite{silberberg_2ph_nonres1_2,baumert_2ph_nonres} and
resonance-mediated (1+1) two-photon absorption
\cite{silberberg_2ph_1plus1,girard_2ph_1plus1,becker_2ph_1plus1_theo}
in atoms, over non-resonant two-photon absorption and non-resonant
three-photon absorption in large molecules
\cite{dantus_2ph_nonres_molec1_2}, over molecular Raman transitions
\cite{leone_res_nonres_raman_control}, and over molecular coherent
anti-Stokes Raman scattering (CARS)
\cite{silberberg_antiStokes_Raman_spect,leone_cars}.

In addition to these multiphoton control works, in a recent study we
firstly analyzed theoretically and demonstrated experimentally
coherent phase control over the process of atomic resonance-mediated
(2+1) three-photon absorption \cite{amitay_3ph_2plus1}. The present
work is a direct continuation of this first work. It extends the
analysis of the corresponding interference mechanism and obtains new
physical insight regarding its dependence on the spectral bandwidth
of the excitation pulse.
The model system of the study is atomic sodium (Na).
Section~\ref{sec:theory} of the paper gives the corresponding
frequency-domain theoretical description of the process as obtained
based on 3$^{rd}$-order time-dependent perturbation theory. The
paper ends with conclusions in Sec.~\ref{sec:conclusions}.
Section~\ref{sec:model-sys} describes the Na model system and the
families of shaped pulses that serve as test cases. It follows by
Sec.~\ref{sec:exp_results} that presents corresponding experimental
results that re-confirm the theoretical description and
corresponding numerical calculations. Then,
Sec.~\ref{sec:theo_results} presents extended numerical-theoretical
results that fully reveal the spectral-bandwidth dependence of the
process, as discussed in Sec.~\ref{sec:discussion}.

\section{\label{sec:theory}Frequency-domain theoretical description}

\begin{figure} [thbp]
\includegraphics[scale=0.5]{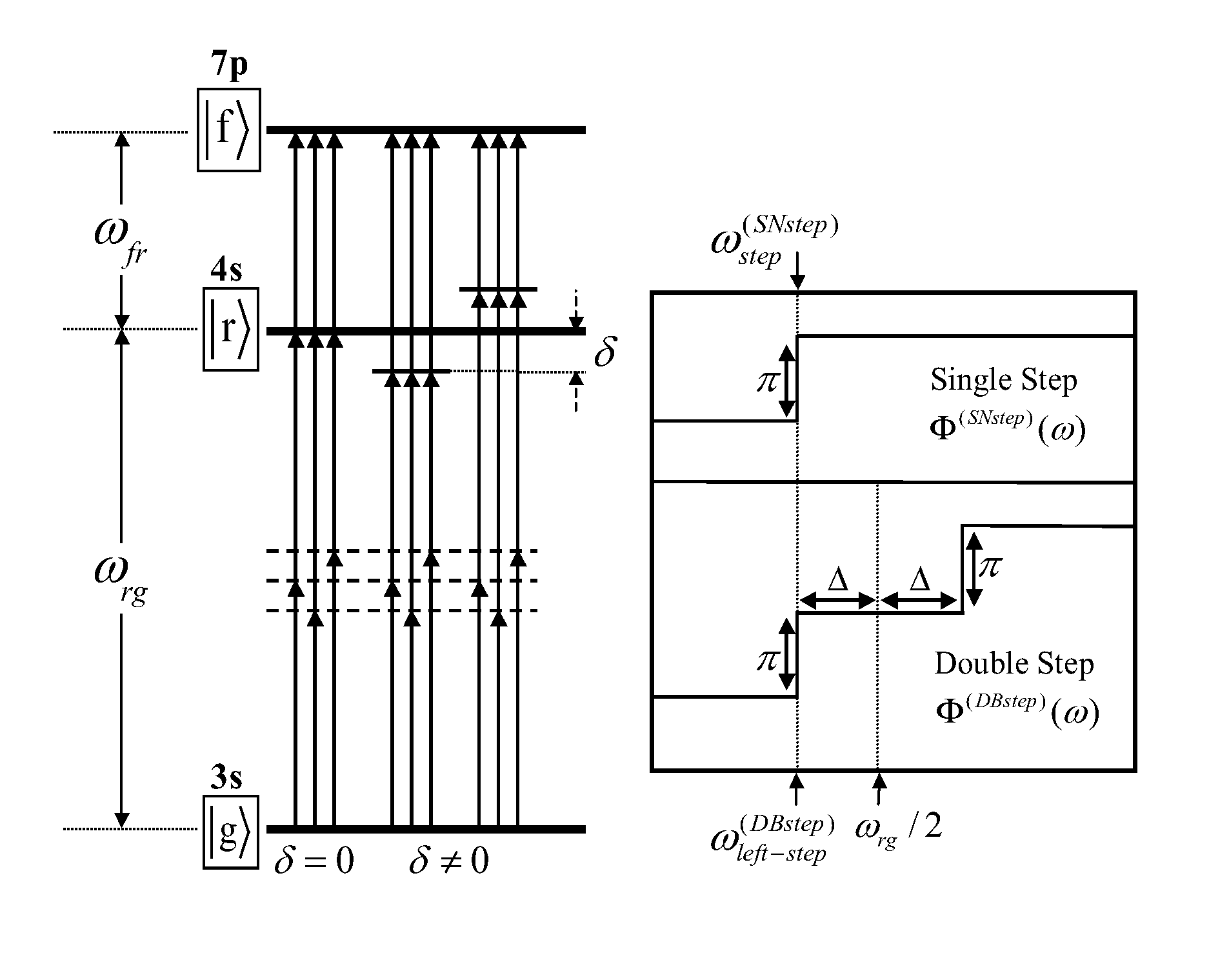}
\caption{\label{fig1} The resonance-mediated (2+1) three-photon
excitation scheme (not to scale) from $\left| g \right\rangle$
to $\left| f \right\rangle$ via $\left| r \right\rangle$.
The corresponding Na states are, respectively, $3s$, $4s$, and $7p$.
Several sets of pathways that are on resonance ($\delta=0$)
and near resonance ($\delta \ne 0$) with $\left| r \right\rangle \equiv 4s$ are shown;
$\delta$ is the corresponding detuning.
Also shown schematically are examples of the two types of
spectral phase patterns used in the study,
$\Phi^{\textrm{(SNstep)}}(\omega)$ and
$\Phi^{\textrm{(DBstep)}}(\omega)$.
The $\Phi^{\textrm{(SNstep)}}(\omega)$ set includes the phase
patterns with a single $\pi$ step, each is characterized by the $\pi$
step position $\omega^{\textrm{(SNstep)}}_{\textrm{step}}$. The
$\Phi^{\textrm{(DBstep)}}(\omega)$ set includes the phase patterns
composed of two $\pi$ steps, with the two steps located at equal
distance along the frequency axis from both sides of half the
two-photon transition frequency $\omega_{rg}/2$. Each such double-step pattern is characterized by
the position of the left step
$\omega^{\textrm{(DBstep)}}_{\textrm{left-step}}$.}
\end{figure}

As shown in Fig.~\ref{fig1},
the resonance-mediated (2+1) three-photon absorption process
considered here involves an initial state
$\left|g\right\rangle$, a final state $\left|f\right\rangle$, and
an intermediate state $\left|r\right\rangle$.
The $\left|g\right\rangle$ and $\left|r\right\rangle$ states are
coupled by a non-resonant two-photon coupling provided by a
manifold of states $\left|v\right\rangle$ that are far from resonance,
i.e., half the two-photon transition frequency $\omega_{rg}/2$ falls within
the pulse spectrum while the
$\left|g\right\rangle$-$\left|v\right\rangle$ transition frequencies $\omega_{vg}$
fall outside the pulse spectrum.
The $\left|f\right\rangle$-$\left|r\right\rangle$ coupling is a resonant one-photon
coupling, i.e., the corresponding transition frequency $\omega_{fr}$ falls within the
pulse spectrum.

In a weak-field regime the three-photon absorption is validly
described by third-order time-dependent perturbation theory.
Accordingly, the time-dependent (complex) amplitude $a_{f}(t)$ of state $\left|f\right\rangle$ at time $t$,
due to the interaction with the shaped temporal electric field $\varepsilon(t)$,
is given by
\begin{eqnarray}
\label{eq_af_t}
 a_{f}(t) & & =-\frac{1}{i\hbar^{3}}\sum_{v}
   \mu_{fr}\mu_{rv}\mu_{vg}
   \int_{-\infty}^{t}\int_{-\infty}^{t_1}\int_{-\infty}^{t_2}
    \varepsilon(t_1)\varepsilon(t_2)\varepsilon(t_3)\nonumber\\
    & &\times
    \exp(i\omega_{fr}t_1)\exp(i\omega_{rv}t_2)
    \exp(i\omega_{vg}t_3)
    dt_1dt_2dt_3 ,
\end{eqnarray}
where $\mu_{mn} = \left\langle m \right| \mu \left| n \right\rangle$
and $\omega_{mn} = (E_{m} - E_{n})/\hbar$ are, respectively, the
dipole matrix element and the transition frequency between a pair of
states $\left|m\right\rangle$ and $\left|n\right\rangle$.
The transition frequencies $\omega_{rg}$, $\omega_{fr}$, and $\omega_{fg}$ satisfy the
relation $\omega_{fg} = \omega_{fr} + \omega_{rg}$.

Within the frequency-domain framework, the spectral field of the pulse
$E(\omega) \equiv \left|E(\omega)\right| \exp \left[ i\Phi(\omega) \right]$ is given as
the Fourier transform of $\varepsilon(t)$,
with $\left|E(\omega)\right|$ and $\Phi(\omega)$ being, respectively,
the spectral amplitude and phase of frequency $\omega$.
For the unshaped transform-limited (TL) pulse, $\Phi(\omega)=0$ for any $\omega$.
We also define the normalized spectral field $\widetilde{E}(\omega) \equiv E(\omega) / \left|E_{0}\right| \equiv
\left|\widetilde{E}(\omega)\right| \exp \left[ i\Phi(\omega)
\right]$ that represents the pulse shape, where $\left|E_{0}\right|$ is the peak spectral amplitude.
This allows to clearly distinguish in the expressions given below between the dependence on the pulse intensity
and the dependence on the pulse shape.
The maximal spectral intensity $I_{0}$ is proportional to $|E_{0}|^{2}$ ($I_{0}\propto|E_{0}|^{2}$).
The resonance-mediated (2+1) three-photon excitation scheme implies that
$\left|\widetilde{E}(\omega_{rg}/2)\right| \ne 0$,
$\left|\widetilde{E}(\omega_{vg})\right| = 0$, and
$\left|\widetilde{E}(\omega_{fr})\right| \ne 0$.
Then, the final amplitude $A_{f}^{(2+1)} \equiv a_f(t \rightarrow
\infty)$ of state $\left| f \right\rangle$ after the pulse is over ($t \rightarrow \infty$)
is obtained to be \cite{amitay_3ph_2plus1}
\begin{equation}
A_f^{(2+1)} = \frac{1}{\hbar^3} \mu_{fr} \mu_{rg}^2 \left|E_{0}\right|^{3}
\left[ A_f^{(2+1)on-res} + A_f^{(2+1)near-res} \right] ,
\label{eq_Af}
\end{equation}
where $\mu_{fr}$ is the $\left|f\right\rangle$-$\left|r\right\rangle$ dipole matrix element
and $\mu_{rg}^2$ is the $\left|r\right\rangle$-$\left|g\right\rangle$ effective non-resonant
two-photon dipole coupling \cite{silberberg_2ph_nonres1_2}.
The terms $A_f^{(2+1)on-res}$ and $A_f^{(2+1)near-res}$ are given by
\begin{eqnarray}
A_f^{(2+1)on-res} & = & 
i\pi \widetilde{E}(\omega_{fr})A^{(2)}(\omega_{rg}) \; ,
\label{eq_Af_on_res}
\\
A_f^{(2+1)near-res} & = &
-\wp\int_{-\infty}^{\infty} \frac{1}{\delta} A^{(2)}(\omega_{rg}-\delta) \widetilde{E}(\omega_{fr}+\delta) d\delta \; ,
\label{eq_Af_near_res}
\end{eqnarray}
with
\begin{eqnarray}
A^{(2)}(\Omega) & = & \int_{-\infty}^{\infty}
\widetilde{E}(\omega)\widetilde{E}(\Omega-\omega)d\omega \; .
\label{eq_A2}
\end{eqnarray}
The expressions of Eqs.~(\ref{eq_Af})-(\ref{eq_A2}) are
equivalent to the ones given in Ref.~\cite{amitay_3ph_2plus1}.
However, as shown below, they are given here in a different form that
provides deeper and more intuitive physical insight to the femtosecond
excitation process.

The final amplitude $A_{f}^{(2+1)}$ of $\left| f \right\rangle$
coherently interferes all the possible three-photon pathways from
$\left|g\right\rangle$ to $\left|f\right\rangle$, i.e., coherently
integrates over all their corresponding amplitudes.
Several pathway examples are shown schematically in Fig.~\ref{fig1}.
Each three-photon pathway is either on-resonance or near-resonance with the intermediate
state $\left|r\right\rangle$, with a corresponding detuning $\delta$.
A resonance-mediated (2+1) three-photon pathway corresponding to a detuning $\delta$
involves a non-resonant absorption of two photons to a total excitation energy of $\omega_{rg}-\delta$
and the absorption of a third photon of frequency $\omega_{fr}+\delta$.
The term $A_{f}^{(2+1)on-res}$ interferes all the on-resonant pathways ($\delta=0$),
while the term $A_{f}^{(2+1)near-res}$ interferes all the near-resonant pathways ($\delta\ne0$)
with an amplitude weighting of $1/\delta$ (i.e., inversely proportional to its detuning).
The on-resonant pathways are excluded from $A_{f}^{(2+1)near-res}$ by
the Cauchy's principal value operator $\wp$.
Thus, the resulting absorption is determined by both intra-group and inter-group interferences
involving these two groups of three-photon excitation pathways.
Both $A_{f}^{(2+1)on-res}$ and $A_{f}^{(2+1)near-res}$ terms are expressed
using the term $A^{(2)}(\Omega)$ that is proportional 
to the non-resonant two-photon absorption amplitude to a total transition frequency of $\Omega$.
It interferes all the corresponding two-photon pathways.
The phase associated with each two-photon pathway is equal to the sum of the phases of the two corresponding photons.
So, with the transform-limited (TL) pulse all the two-photon pathways acquire zero phase for any value of $\Omega$
and, thus, interfere one with the other in a fully constructive way.
For a given field spectrum $\left|\widetilde{E}(\omega)\right|$, this leads to the maximal $A^{(2)}(\Omega)$ for any value of $\Omega$.
For a given $\Omega$, any spectral phase pattern that is anti-symmetric around $\Omega/2$ also
induces fully constructive interferences among all the two-photon pathways of $\Omega$ transition frequency,
and thus it induces the corresponding maximal $A^{(2)}(\Omega)$ as the TL pulse.
As can be seen from Eq.~(\ref{eq_A2}), $A^{(2)}(\Omega)$ is actually the
(complex) convolution function of the (complex) field $\widetilde{E}(\omega)$
evaluated at $\Omega$.
Since $A_f^{(2+1)on-res}$ is proportional to $A^{(2)}(\omega_{rg})$,
it is actually proportional to the weak-field non-resonant two-photon absorption
amplitude to $\left|r\right\rangle$, as
obtained by second-order perturbation theory \cite{silberberg_2ph_nonres1_2}.

The final population $P_{f}^{(2+1)}$ of state $\left|f\right\rangle$,
given by $P_{f}^{(2+1)} = \left|A_f^{(2+1)}\right|^{2}$,
serves as the measure for the resonance-mediated (2+1) three-photon absorption.
All the results presented here are given relative to the absorption induced by the
transform-limited (TL) pulse. So, for convenience, we introduce the
TL-normalized three-photon absorption measure
$\widetilde{P}_{f}^{(2+1)} = P_{f}^{(2+1)} / P_{f,\textrm{TL}}^{(2+1)}$,
where $P_{f,\textrm{TL}}^{(2+1)}$ is the final $\left|f\right\rangle$ population
induced by the TL pulse.

\section{\label{sec:results}Results}

\subsection{\label{sec:model-sys}Model system and pulse-shape test cases}
The model system of the study is the sodium (Na) atom \cite{NIST},
with the corresponding excitation scheme shown in Fig.~\ref{fig1}.
It involves the $3s$ ground state as $\left|g\right\rangle$, the
$4s$ state as $\left|r\right\rangle$, and the $7p$ state as
$\left|f\right\rangle$.
The transition frequency $\omega_{rg} \equiv \omega_{4s,3s} =
25740$~cm$^{-1}$ corresponds to two 777~nm photons and the
transition frequency $\omega_{fr} \equiv \omega_{7p,4s} =
12801$~cm$^{-1}$ corresponds to a one 781.2~nm photon.
The $3s$-$4s$ non-resonant two-photon coupling originates from
the manifold of $p$-states,
particularly from the $3p$ state [$\omega_{3p,3s} \sim
16978$~cm$^{-1}$ (589~nm)].
The absorption measure is the final $7p$ population $P_{f}^{(2+1)} \equiv P_{7p}^{(2+1)}$.

As the test cases for coherent phase control, the present study uses
shaped femtosecond pulses having two sets of spectral phase
patterns, $\Phi^{\textrm{(SNstep)}}(\omega)$ and
$\Phi^{\textrm{(DBstep)}}(\omega)$.
They are shown schematically in the inset of Fig.~\ref{fig1}.
The $\Phi^{\textrm{(SNstep)}}(\omega)$ set includes the phase patterns of 
a single $\pi$ step, each characterized by the $\pi$ step position $\omega^{\textrm{(SNstep)}}_{\textrm{step}}$.
The $\Phi^{\textrm{(DBstep)}}(\omega)$ set includes the phase patterns composed of
two $\pi$ steps, with the two steps located at equal distance along the frequency axis
from both sides (i.e., to the blue and to the red) of half the two-photon transition frequency $\omega_{4s,3s}/2$,
i.e., $\omega_{4s,3s}/2-\omega^{\textrm{(DBstep)}}_{\textrm{left-step}}=
\omega^{\textrm{(DBstep)}}_{\textrm{right-step}}-\omega_{4s,3s}/2 \equiv \Delta^{\textrm{(DBstep)}}_{\textrm{step}}$.
In our previous work \cite{amitay_3ph_2plus1} the single-step phase
patterns have been shown to be very effective in controlling the
resonance-mediated (2+1) three-photon absorption.
The double-step phase patterns are additionally included here since, as explained below,
they all induce the same maximal on-resonant component $A_{7p}^{(2+1)on-res}$, while each of them
induces a different near-resonant component $A_{7p}^{(2+1)near-res}$.

The $\pi$ amplitude of all the phase steps, i.e., generally
$\Phi(\omega)=$ 0 or $\pi$, implies that the field
$\widetilde{E}(\omega)$ is always a real (positive or negative)
quantity for any $\omega$. Thus, the corresponding
$A_{7p}^{(2+1)on-res}$ and $A_{7p}^{(2+1)near-res}$ are,
respectively, purely imaginary and real [see
Eqs.~(\ref{eq_Af_on_res})-(\ref{eq_Af_near_res})],
and $P_{7p}^{(2+1)} \propto \left|A_{7p}^{(2+1)on-res}\right|^{2} +
\left|A_{7p}^{(2+1)near-res}\right|^{2}$. Consequently,
$\widetilde{P}_{7p}^{(2+1)} =
\left|\widetilde{A}_{7p}^{(2+1)on-res}\right|^{2} +
\left|\widetilde{A}_{7p}^{(2+1)near-res}\right|^{2}$, where
$\left|\widetilde{A}_{7p}^{(2+1)on-res}\right|^{2} =
 \left|A_{7p}^{(2+1)on-res}\right|^{2} / P_{7p,\textrm{TL}}^{(2+1)}$ 
and 
$\left|\widetilde{A}_{7p}^{(2+1)near-res}\right|^{2} =
 \left|A_{7p}^{(2+1)near-res}\right|^{2} / P_{7p,\textrm{TL}}^{(2+1)}$. 
In other words, with the spectral phase patterns discussed here, the absorption is determined only by intra-group interferences
taking place separately within each of the on-resonant and near-resonant excitation groups,
without any inter-group interferences taking place.
This fact greatly simplify, without reduction of generality, the ability to gain physical insight into
the pulse-bandwidth dependence of the resonance-mediated (2+1) three-photon absorption process.
%


\subsection{\label{sec:exp_results}Experimental vs. numerical-theoretical results}

\begin{figure} [htbp]
\includegraphics[scale=0.5]{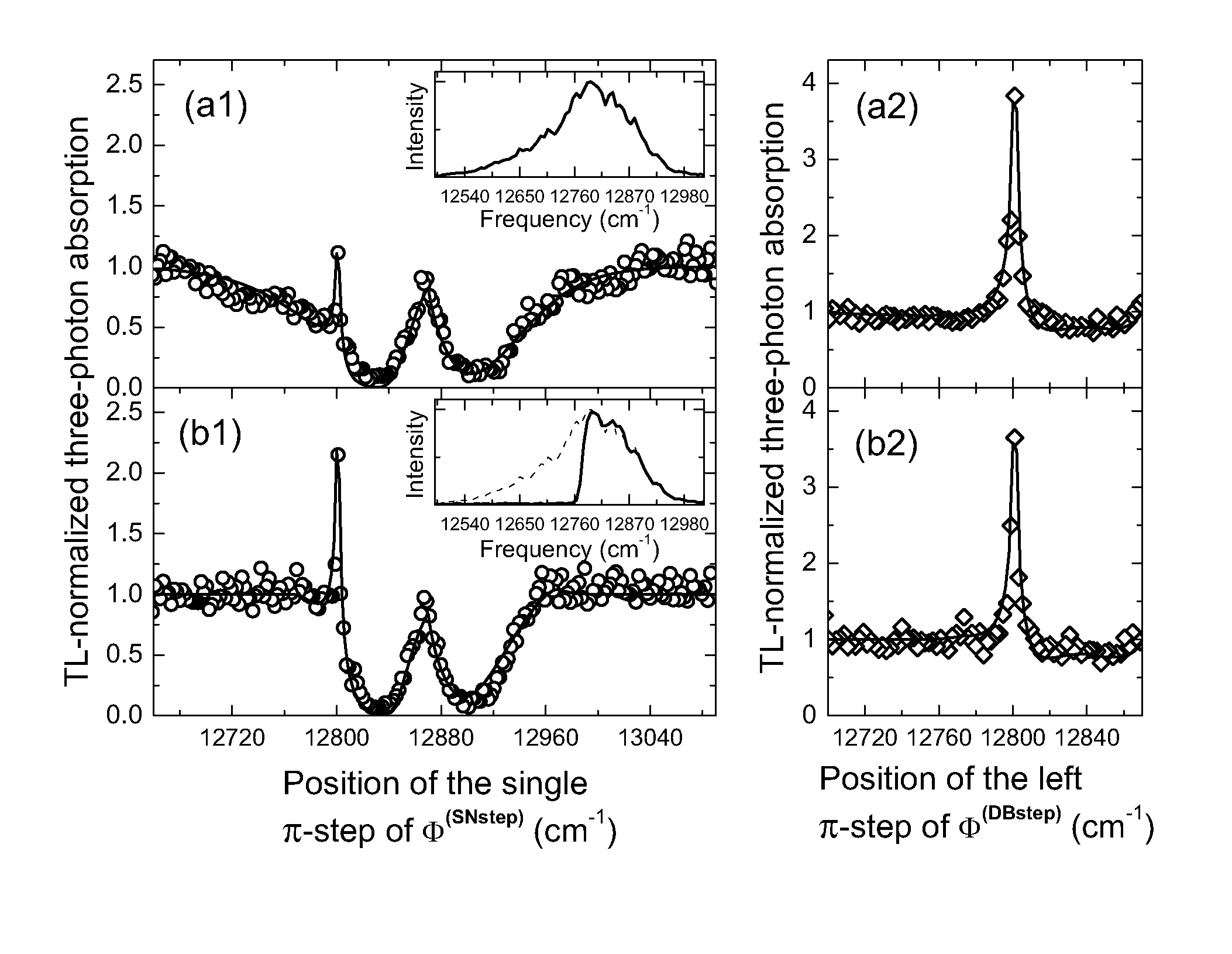}
\caption{\label{fig2} Experimental (circles and squares) and
numerical-theoretical (solid lines) results for the
resonance-mediated (2+1) three-photon absorption in Na.
Panels (a1) and (a2) correspond to the full pulse spectrum [inset of (a1)],
while panels (b1) and (b2) correspond to the blocked pulse spectrum [inset of (b1)].
Panels (a1) and (b1) (left-column panels) present results for the
single-step set of spectral phase patterns $\Phi^{\textrm{(SNstep)}}(\omega)$.
The corresponding traces show the TL-normalized final $7p$
population $\widetilde{P}_{7p}^{(2+1)}$ (see text) as a function of
the single $\pi$-step position
$\omega^{\textrm{(SNstep)}}_{\textrm{step}}$.
Panels (a2) and (b2) (right-column panels) present results for the
double-step set of spectral phase patterns $\Phi^{\textrm{(DBstep)}}(\omega)$.
The corresponding traces show the TL-normalized final $7p$
population $\widetilde{P}_{7p}^{(2+1)}$ (see text) as a function of
the left $\pi$-step position
$\omega^{\textrm{(DBstep)}}_{\textrm{left-step}}$.}
\end{figure}

This section presents experimental results for the Na atom, re-confirming the above theoretical description
with spectral parameters, i.e., spectral bandwidth and the above phase patterns, that are different
from those used in our previous work \cite{amitay_3ph_2plus1}.
Experimentally, atomic sodium vapor is produced in a static chamber at
$300^{o}$C (Na partial pressure of $\sim$0.1 Torr) with 10-Torr Ar buffer gas.
It is irradiated at a 1-kHz repetition rate with linearly-polarized shaped femtosecond laser pulses
of 9.8-nm (160-cm$^{-1}$) intensity bandwidth (FWHM) around 781~nm (12804~cm$^{-1}$).
The spectrum shape is a modified Gaussian with a slight asymmetry toward low frequencies.
It is shown in the inset of Fig.~\ref{fig2}(a1).
Experiments are also conducted with a spectrum that results from blocking about half of this full spectrum
at a cutoff wavelength of 783~nm (12771~cm$^{-1}$).
It is shown in the inset of Fig.~\ref{fig2}(b1).
In both spectral cases, the temporal peak intensity of the corresponding transform-limited (TL) pulse 
is below 10$^{9}$~W/cm$^2$.
The laser pulses undergo shaping in a 4$f$ optical setup incorporating a pixelated
liquid-crystal spatial light phase modulator \cite{pulse_shaping}.
The effective spectral shaping resolution is $\delta\omega_{shaping}=2.05$~cm$^{-1}$ (0.125~nm) per pixel.
Following the interaction with a pulse, the Na population excited to the $7p$ state undergoes
radiative and collisional decay to lower excited states, including the $4d$, $5d$, $6d$, and $6s$ states.
The fluorescence emitted in their decay to the $3p$ state 
serves as the relative measure for the total final $7p$ population $P_{7p}^{(2+1)}$.
It is optically measured at 90$^{\circ}$ to the beam propagation direction
using a spectrometer coupled to a time-gated camera system.

Figure~\ref{fig2} compares the experimental results (circles or squares)
with corresponding theoretical results (solid lines)
for the resonance-mediated (2+1) three-photon absorption in Na.
The theoretical results are calculated numerically using Eqs.~(\ref{eq_Af})-(\ref{eq_A2}),
using a grid with a bin size equal to the experimental spectral shaping resolution $\delta\omega_{shaping}$.
Figures~\ref{fig2}(a1) and \ref{fig2}(a2) (first-row panels) correspond to the full pulse spectrum [inset of Fig.~\ref{fig2}(a1)],
while Figs.~\ref{fig2}(b1) and \ref{fig2}(b2) (second-row panels) correspond to the blocked pulse spectrum [inset of Fig.~\ref{fig2}(b1)].
Figures~\ref{fig2}(a1) and \ref{fig2}(b1) (left-column panels) present results for
the single-step set of spectral phase patterns $\Phi^{\textrm{(SNstep)}}(\omega)$.
The corresponding traces show the TL-normalized final $7p$ population $\widetilde{P}_{7p}^{(2+1)}$ (see above)
as a function of the single $\pi$-step position $\omega^{\textrm{(SNstep)}}_{\textrm{step}}$.
Figures~\ref{fig2}(a2) and \ref{fig2}(b2) (right-column panels) present results for
the double-step set of spectral phase patterns $\Phi^{\textrm{(DBstep)}}(\omega)$.
The corresponding traces show the TL-normalized final $7p$ population $\widetilde{P}_{7p}^{(2+1)}$ (see above)
as a function of the left $\pi$-step position $\omega^{\textrm{(DBstep)}}_{\textrm{left-step}}$.

As can be seen, 
there is an excellent quantitative agreement between the
experimental results ("real experiment") and the
numerical-theoretical results ("computer experiment"). Hence,
supported also by our previous work \cite{amitay_3ph_2plus1}, the
validity and accuracy of the theoretical description and numerical
calculations are re-confirmed.
As such, in the following they serve us for the detailed study of the $\pi$-traces,
their spectral-bandwidth dependence, and the corresponding interference mechanism.
%


\subsection{\label{sec:theo_results}Numerical-theoretical results}

\begin{figure} [htbp]
\includegraphics[scale=0.5]{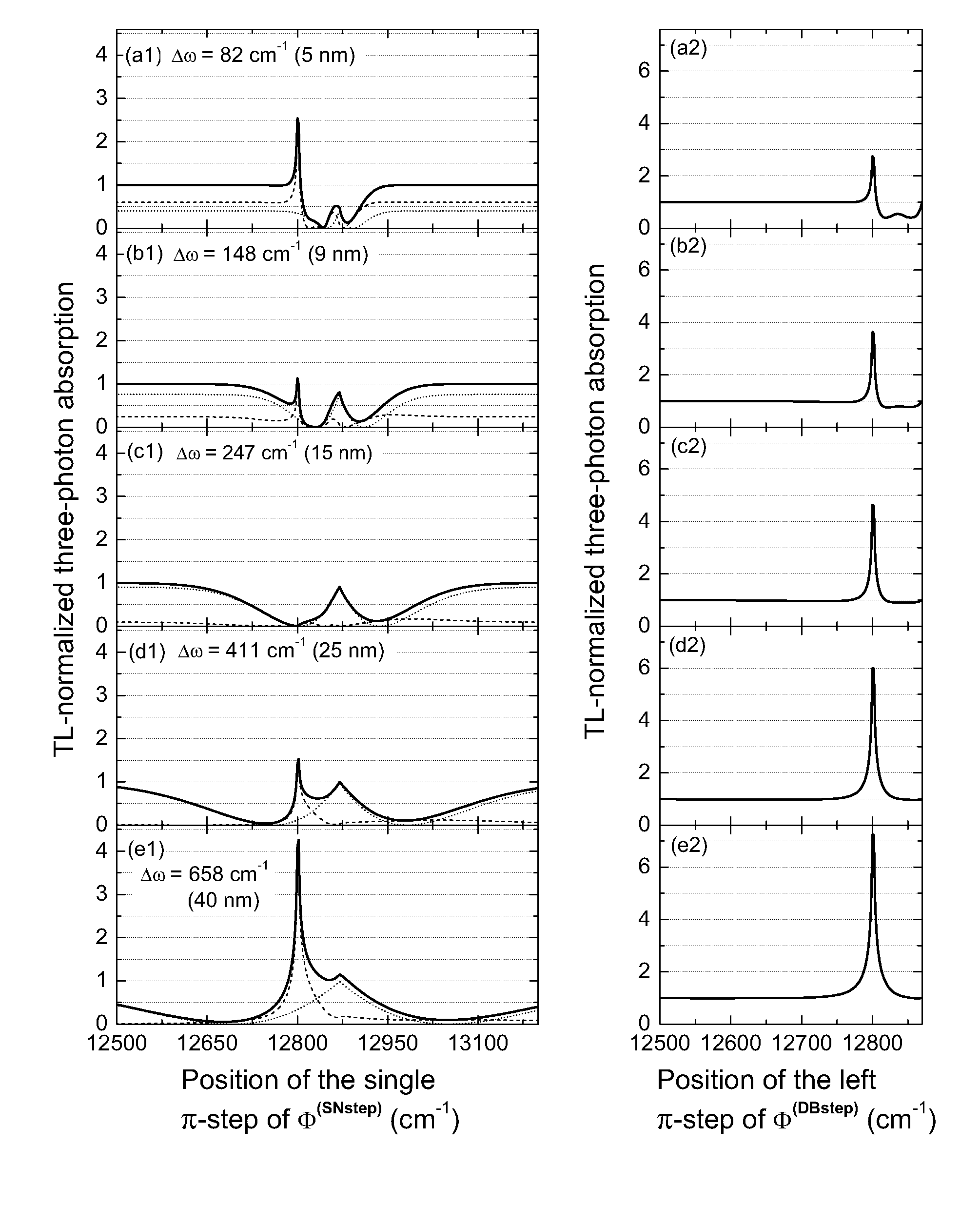}
\caption{\label{fig3} Theoretical results (solid lines) of the
resonance-mediated (2+1) three-photon absorption in Na calculated
with excitation pulse of different spectral intensity bandwidth
$\Delta\omega$.
In all the cases the pulse spectrum is a Gaussian centered at 780~nm (12821~cm$^{-1}$).
Each row in the figure corresponds to a different bandwidth.
The left-column panels correspond to the single-step spectral phase
patterns $\Phi^{\textrm{(SNstep)}}(\omega)$. They show the TL-normalized final $7p$ population
$\widetilde{P}_{7p}^{(2+1)}$ (solid lines) as a function of the single $\pi$-step
position $\omega^{\textrm{(SNstep)}}_{\textrm{step}}$. Additionally,
they also show the corresponding on-resonant component
$\left|\widetilde{A}_{7p}^{(2+1)on-res}\right|^{2}$ (dotted lines)
and near-resonant component
$\left|\widetilde{A}_{7p}^{(2+1)near-res}\right|^{2}$ (dashed
lines).
The right-column panels correspond to the double-step spectral phase
patterns $\Phi^{\textrm{(DBstep)}}(\omega)$. They show the
TL-normalized final $7p$ population $\widetilde{P}_{7p}^{(2+1)}$ as
a function of the left $\pi$-step position
$\omega^{\textrm{(DBstep)}}_{\textrm{left-step}}$.}
\end{figure}

Figure~\ref{fig3} presents theoretical results (solid lines) that have been calculated numerically
for the resonance-mediated (2+1) three-photon absorption in Na with excitation pulse of
different spectral bandwidth.
In all the cases the intensity spectrum is a Gaussian centered at 780~nm (12821~cm$^{-1}$).
The corresponding bandwidth (FWHM) values ($\Delta\omega$)
are (a) 5~nm (82~cm$^{-1}$), (b) 9~nm (148~cm$^{-1}$), (c) 15~nm (247~cm$^{-1}$),
(d) 25~nm (411~cm$^{-1}$), and (e) 40~nm (658~cm$^{-1}$).
Each row in the figure corresponds to a different bandwidth.
As indicated above, the numerical calculations are based on Eqs.~(\ref{eq_Af})-(\ref{eq_A2}) and
use a grid with a bin size equal to the experimental spectral shaping resolution.

The left-column panels of Fig.~\ref{fig3} correspond to the single-step spectral phase patterns $\Phi^{\textrm{(SNstep)}}(\omega)$.
In solid lines they show the TL-normalized final $7p$ population $\widetilde{P}_{7p}^{(2+1)}$
as a function of the single $\pi$-step position $\omega^{\textrm{(SNstep)}}_{\textrm{step}}$.
Additionally, they also show the corresponding on-resonant component $\left|\widetilde{A}_{7p}^{(2+1)on-res}\right|^{2}$ (dotted lines)
and the near-resonant component $\left|\widetilde{A}_{7p}^{(2+1)near-res}\right|^{2}$ (dashed lines),
satisfying $\widetilde{P}_{7p}^{(2+1)} = \left|\widetilde{A}_{7p}^{(2+1)on-res}\right|^{2} + \left|\widetilde{A}_{7p}^{(2+1)near-res}\right|^{2}$
(see above).
As can be seen, the shape of the traces strongly depends on the spectral bandwidth.
For any given bandwidth, the three-photon absorption is tunable over a wide range of values,
from a low attenuated level below 0.01 of the TL absorption to a
high enhanced level with a value that strongly depends on the spectral bandwidth.
Whenever this enhanced absorption exceeds the TL absorption, it
occurs when $\omega^{\textrm{(SNstep)}}_{\textrm{step}} =
\omega_{7p,4s}$, i.e., when the $\pi$-step is located at the
$7p$-$4s$ transition frequency.
This is actually the location of the prominent feature of the single-step traces.
As seen in Fig.~\ref{fig3}, as the spectral bandwidth increases from
5~nm to 9~nm to 15~nm, the value of $\widetilde{P}_{7p}^{(2+1)}$ at
$\omega^{\textrm{(SNstep)}}_{\textrm{step}}=\omega_{7p,4s}$
decreases, respectively, from 2.5 to 1.1 to below 0.01. Then,
further increase in the bandwidth leads to its increase to a value
of 1.4 at 25-nm bandwidth and to a value of 4.1 at 40-nm bandwidth.
Another prominent bandwidth dependence is observed for the TL pulse regarding the relative weight of the
corresponding on-resonant and near-resonant amplitude components,
as it is reflected in the asymptotes of the single-step traces:
As the bandwidth increases, the TL on-resonant component dominates more and more the total absorption,
approaching the limit of $\left|A_{7p,\textrm{TL}}^{(2+1)near-res}\right| \ll \left|A_{7p,\textrm{TL}}^{(2+1)on-res}\right|$
and $\left|\widetilde{A}_{7p,\textrm{TL}}^{(2+1)on-res}\right|^{2} \rightarrow 1$.

The right-column panels of Fig.~\ref{fig3} correspond to the double-step spectral phase patterns $\Phi^{\textrm{(DBstep)}}(\omega)$.
They show the TL-normalized final $7p$ population $\widetilde{P}_{7p}^{(2+1)}$
as a function of the left $\pi$-step position $\omega^{\textrm{(DBstep)}}_{\textrm{left-step}}$.
One should note that the maximal value of the y-axis in these traces is higher than in the traces
of the single-step phase patterns (the left-column panels).
The prominent feature of the double-step traces is the strong
absorption enhancement occurring when
$\omega^{\textrm{(DBstep)}}_{\textrm{left-step}} = \omega_{7p,4s}$.
This is also the step-position region, where the dependence on the
spectral bandwidth is the strongest.
The corresponding enhancement value is always above one, i.e., above the TL absorption,
and it continuously increases as the bandwidth increases.
It starts from a value of 2.8 at 5-nm bandwidth [Fig.~\ref{fig3}(a2)]
and reaches a value of 7.2 at 40-nm bandwidth [Fig.~\ref{fig3}(e2)].
When $\omega^{\textrm{(DBstep)}}_{\textrm{left-step}}$ is far from
$\omega_{7p,4s}$, the three-photon absorption is always kept on a
level that is very close to or equal to the TL absorption. The only
exception is the case of 5-nm bandwidth [Fig.~\ref{fig3}(a2)], where
at step positions higher than $\omega_{7p,4s}$ the absorption is
reduced to a level of about half the TL absorption.

\begin{figure} [htbp]
\includegraphics[scale=0.5]{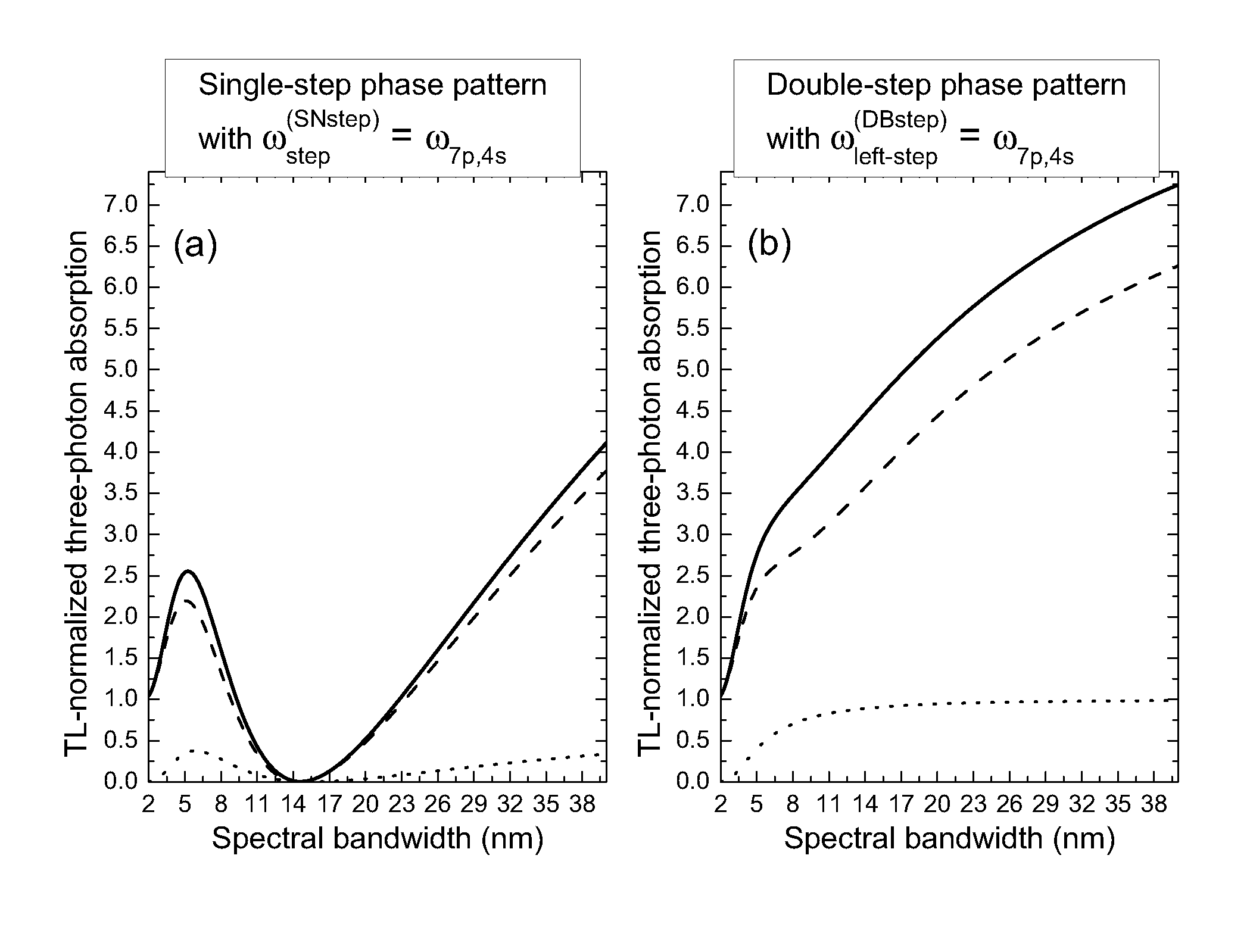}
\caption{\label{fig4} Numerical-theoretical results for the
dependence of the resonance-mediated (2+1) three-photon absorption
in Na on the spectral intensity bandwidth of the excitation pulse
having
(a) single-step phase pattern of $\omega^{\textrm{(SNstep)}}_{\textrm{step}} = \omega_{7p,4s}$,
and
(b) double-step phase pattern of $\omega^{\textrm{(DBstep)}}_{\textrm{left-step}} = \omega_{7p,4s}$.
Shown are $\widetilde{P}_{f}^{(2+1)}$ (solid lines),
$\left|\widetilde{A}_{7p}^{(2+1)on-res}\right|^{2}$ (dotted lines),
and $\left|\widetilde{A}_{7p}^{(2+1)near-res}\right|^{2}$ (dashed
lines) as a function of the spectral bandwidth $\Delta\omega$.
}\end{figure}

As can be seen in Fig.~\ref{fig3}, the main features of the traces
when $\omega^{\textrm{(SNstep)}}_{\textrm{step}} = \omega_{7p,4s}$ and
when $\omega^{\textrm{(DBstep)}}_{\textrm{left-step}} = \omega_{7p,4s}$
originate from the near-resonant component $\left|\widetilde{A}_{7p}^{(2+1)near-res}\right|^{2}$.
Thus, to complete the picture, Fig.~\ref{fig4} presents extended results for the spectral-bandwidth  
dependence of the resonance-mediated (2+1) three-photon absorption
when $\omega^{\textrm{(SNstep)}}_{\textrm{step}} = \omega_{7p,4s}$ in a single-step pattern [Fig.~\ref{fig4}(a)]
and when $\omega^{\textrm{(DBstep)}}_{\textrm{left-step}} = \omega_{7p,4s}$ in a double-step pattern [Fig.~\ref{fig4}(b)].
The figure presents the corresponding values of
$\widetilde{P}_{f}^{(2+1)}$ (solid lines),
$\left|\widetilde{A}_{7p}^{(2+1)on-res}\right|^{2}$ (dotted lines),
and $\left|\widetilde{A}_{7p}^{(2+1)near-res}\right|^{2}$ (dashed
lines) as a function of the spectral bandwidth.
The bandwidth values span the wide range of 2 to 40~nm that corresponds, respectively,
to a TL pulse duration of 450 down to 22~fs.
As can be seen, with the single-step phase pattern of
$\omega^{\textrm{(SNstep)}}_{\textrm{step}} = \omega_{7p,4s}$, the
degree of absorption is between 0.01 to 4 times the TL absorption,
i.e., it is either enhanced or attenuated as compared to the TL
case. There is a local peak of value 2.55 at a bandwidth of 5.25~nm,
and a minimum of zero theoretical value (i.e., three-photon dark
pulse) at a bandwidth value of 14.6~nm. Then, above 14.6~nm,
$\widetilde{P}_{f}^{(2+1)}$ is monotonically increases with the
increase in the bandwidth. At 23-nm bandwidth it reaches a value of
1, at 32~nm it reaches again a value of 2.5, and at 40~nm it reaches
a value of 4.1~.
On the other hand, with the double-step phase pattern of $\omega^{\textrm{(DBstep)}}_{\textrm{left-step}} = \omega_{7p,4s}$,
the absorption is always above the TL absorption and the corresponding enhancement monotonically increases with the increase
in the spectral bandwidth.
At 2-nm bandwidth $\widetilde{P}_{7p}^{(2+1)}$ is of a value of 1, and it reaches a value of 7.2 at a bandwidth of 40~nm.
As can be seen in Fig.~\ref{fig4}, in both these single-step and double-step patterns having the step at $\omega_{7p,4s}$,
over the full bandwidth range, the near-resonant component $\left|\widetilde{A}_{7p}^{(2+1)near-res}\right|^{2}$
is much larger than the on-resonant component $\left|\widetilde{A}_{7p}^{(2+1)on-res}\right|^{2}$.


\section{\label{sec:discussion}Analysis and Discussion}

Below, the main features of the results presented above are analyzed based on the theoretical description
formulated in Eqs.~(\ref{eq_Af})-(\ref{eq_A2}).
As noted above, for the spectral phase patterns used here, the
resonance-mediated (2+1) three-photon absorption is determined only
by intra-group interferences within each of the on-resonant and
near-resonant excitation groups that correspond, respectively, to
$A_{7p}^{(2+1)on-res}$ and $A_{7p}^{(2+1)near-res}$, with the
relations $P_{7p}^{(2+1)} \propto
\left|A_{7p}^{(2+1)on-res}\right|^{2} +
\left|A_{7p}^{(2+1)near-res}\right|^{2}$ and
$\widetilde{P}_{7p}^{(2+1)} =
\left|\widetilde{A}_{7p}^{(2+1)on-res}\right|^{2} +
\left|\widetilde{A}_{7p}^{(2+1)near-res}\right|^{2}$.

The analysis uses the numerical results presented in Fig.~\ref{fig5}.
Each panel in the figure corresponds to a pulse having a different spectral bandwidth
for its Gaussian intensity spectrum around 780~nm.
The corresponding bandwidth values are (a) 5~nm, (b) 9~nm, (c) 15~nm, (d) 25~nm, and (e) 40~nm,
as they are in the results shown in Fig.~\ref{fig3}.
Each panel shows results for the transform-limited (TL) pulse (dashed black line),
for the shaped pulse with the single-step phase pattern of
$\omega^{\textrm{(SNstep)}}_{\textrm{step}} = \omega_{7p,4s}$ (solid gray line),
and for the shaped pulse with
the double-step phase pattern of $\omega^{\textrm{(DBstep)}}_{\textrm{left-step}} = \omega_{7p,4s}$
(solid black line).
For the different cases, 
the values of $A^{(2)}(\Omega=\omega_{4s,3s} - \delta)$ [see Eq.~(\ref{eq_A2})]
are shown as a function of the detuning $\delta$ from the $4s$ state.
The detuning (x-axis) values are actually given as normalized values of $\delta/\Delta\omega$,
with $\Delta\omega$ being the bandwidth of the intensity spectrum.

As seen from Eq.~(\ref{eq_Af_on_res}),
the value of the on-resonant component $A_{7p}^{(2+1)on-res}$ is proportional to
$A^{(2+1)}(0) \equiv A^{(2)}(\omega_{4s,3s}) \widetilde{E}(\omega_{7p,4s})$ that corresponds to zero detuning $\delta=0$.
On the other hand, as seen from Eq.~(\ref{eq_Af_near_res}),
due to the $1/\delta$-weighting, the result of the $\wp$-integration of $A_{7p}^{(2+1)near-res}$
is dominated by the integration over small non-zero values of $\left|\delta\right|$ around $\delta$=0.
Due to the sign change of $1/\delta$ for $\pm\left|\delta\right|$, the $\wp$-integration result is 
highly sensitive to the symmetry of the integrand term
$A^{(2+1)}(\delta) \equiv A^{(2)}(\omega_{4s,3s}-\delta) \widetilde{E}(\omega_{7p,4s}+\delta)$ around $\delta$=0,
i.e., its relative magnitude and relative sign for positive and negative detunings of equal magnitude
(i.e., $\pm\left|\delta\right|$).
The zone of small $\left|\delta\right|$ around $\delta$=0
is indicated schematically as dashed area in Fig.~\ref{fig5}.

\subsection{Transform-limited pulse}

First we analyze the case of the transform-limited (TL) pulse (dashed thin black lines in Fig.~\ref{fig5}).
Since $A^{(2)}(\Omega)$ is the convolution function of $\widetilde{E}(\omega)$ evaluated at $\Omega$,
for the TL pulse [$\Phi(\omega)=0$] it is actually a convolution of 
$\left|\widetilde{E}(\omega)\right|$ with a real positive values for any $\Omega$.
Thus, with a Gaussian spectrum around $\omega_{0}$, $A_{\textrm{TL}}^{(2)}(\Omega)$ has a Gaussian shape that is
peaked at $\Omega_{peak}=2\omega_{0}$. The corresponding detuning is
$\delta_{peak} = \omega_{4s,3s} - 2 \omega_{0}$. 
The width (FWHM) of the $A_{\textrm{TL}}^{(2)}(\Omega)$ Gaussian profile is $2 \Delta\omega$, where $\Delta\omega$ is the
width of the intensity spectrum.   
Since the detuning values given in Fig.~\ref{fig5} are normalized by $\Delta\omega$,
all the presented TL profiles of $A_{\textrm{TL}}^{(2)}(\Omega=\omega_{4s,3s}-\delta)$ have the same width (of value 2)
for any spectral bandwidth $\Delta\omega$.
On the other hand, an increase in $\Delta\omega$ leads to a decrease in the value of $\delta_{peak}/\Delta\omega$. 
Hence, as seen in Fig.~\ref{fig5}, as $\Delta\omega$ increases,
the peak and the full 
profile of $A_{\textrm{TL}}^{(2)}(\omega_{4s,3s}-\delta)$ shifts to smaller values of $\delta/\Delta\omega$
and the corresponding value of $A_{\textrm{TL}}^{(2)}(\omega_{4s,3s})$,  
which determines the on-resonant component $A_{7p,\textrm{TL}}^{(2+1)on-res}$, increases.

For the TL pulse,
both $A^{(2)}_{\textrm{TL}}(\omega_{4s,3s}-\delta)$ and $\widetilde{E}_{\textrm{TL}}(\omega_{7p,4s}+\delta)$ are
real positive quantities for any $\delta$.
Thus, since the sign of $1/\delta$ is different for +$\left|\delta\right|$ and $-$$\left|\delta\right|$,
the amplitudes $\frac{1}{\delta} A^{(2+1)}_{\textrm{TL}}(\delta)$
contributed by $\pm\left|\delta\right|$ to the $\wp$-integration of $A_{7p,\textrm{TL}}^{(2+1)near-res}$
are of different signs, i.e., they interfere  destructively one with the other.

As a result from all the above, an increase of $\Delta\omega$ leads
to a decrease in the value of $\left| \frac{1}{\left|\delta\right|}
 A^{(2+1)}_{\textrm{TL}}(+\left|\delta\right|) - \frac{1}{\left|\delta\right|} A^{(2+1)}_{\textrm{TL}}(-\left|\delta\right|) \right| /
 \left|A^{(2+1)}_{\textrm{TL}}(0)\right|$ for a given small $\left|\delta\right|$.
Consequently, there is a decrease in the relative magnitude between
the TL near-resonant component $A_{7p,\textrm{TL}}^{(2+1)near-res}$
and the TL on-resonant component $A_{7p,\textrm{TL}}^{(2+1)on-res}$.
Eventually, at large enough $\Delta\omega$, it leads to
$\left|A_{7p,\textrm{TL}}^{(2+1)near-res}\right| \ll \left|A_{7p,\textrm{TL}}^{(2+1)on-res}\right|$,
$P_{7p,\textrm{TL}}^{(2+1)} \rightarrow \left|A_{7p,\textrm{TL}}^{(2+1)on-res}\right|^{2}$, and
$\left|\widetilde{A}_{7p,\textrm{TL}}^{(2+1)on-res}\right|^{2} \rightarrow 1$.
As noted above, this is indeed the behavior seen in the asymptotes of the single-step traces
shown in the left-column panels of Fig.~\ref{fig3} (dashed and dotted lines).
In other words, for the TL pulse, with large enough bandwidth, the absorption amplitudes contributed by
positively-detuned and negatively-detuned near-resonant three-photon pathways completely cancel out each other.
Then, the total TL absorption is determined only by the three-photon pathways that are on-resonant with the intermediate state.

\subsection{Shaped pulse with single-step phase patten of $\omega^{\textrm{(SNstep)}}_{\textrm{step}} = \omega_{7p,4s}$}

\begin{figure} [htbp]
\includegraphics[scale=0.5]{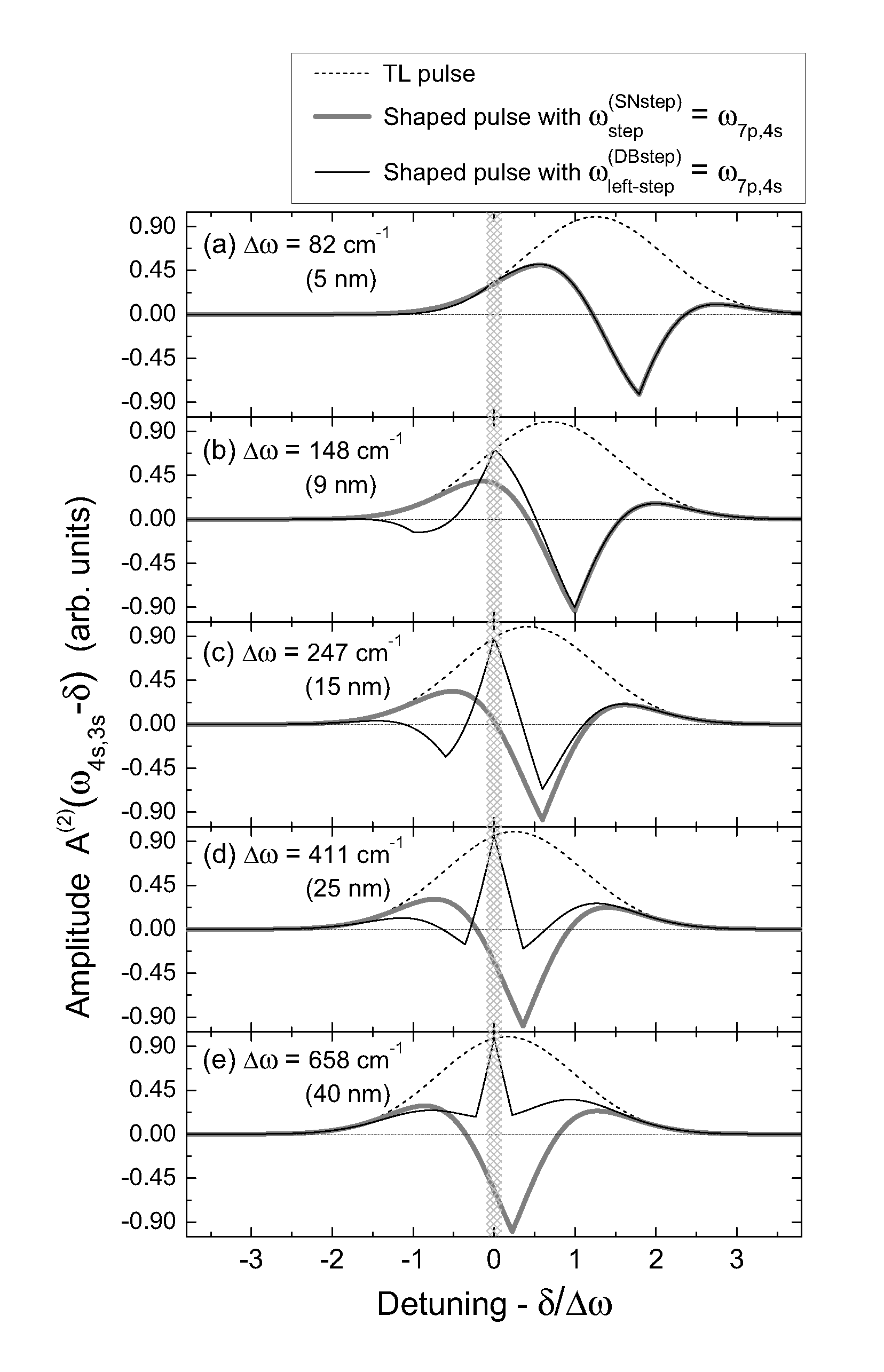}
\caption{\label{fig5} Numerical-theoretical results for the values
of $A^{(2)}(\Omega=\omega_{4s,3s} - \delta)$ [see Eq.~(\ref{eq_A2})]
as a function of the detuning $\delta$ for different spectral
intensity bandwidth $\Delta\omega$ of the excitation pulse.
In all the cases the pulse spectrum is a Gaussian centered at 780~nm (12821~cm$^{-1}$).
The different panels correspond to the different panels in Fig.~\ref{fig3}.
The detuning values are given as normalized values $\delta/\Delta\omega$ (see text).
Each panel shows results for the transform-limited (TL) pulse (dashed black line),
for the shaped pulse with the single-step phase pattern
of $\omega^{\textrm{(SNstep)}}_{\textrm{step}} = \omega_{7p,4s}$ (solid gray line),
and for the shaped pulse with the double-step phase pattern of
$\omega^{\textrm{(DBstep)}}_{\textrm{left-step}} = \omega_{7p,4s}$ (solid black line).}
\end{figure}

Next we consider the case of the shaped pulse with the phase pattern of a single $\pi$ step at
$\omega^{\textrm{(SNstep)}}_{\textrm{step}} = \omega_{7p,4s}$ (solid gray lines in Fig.~\ref{fig5}).
As the corresponding field $\widetilde{E}_{\textrm{SNstep}@\omega_{7p,4s}}(\omega_{7p,4s}+\delta)$ is a real positive or negative
quantity for any $\omega$, the corresponding $A_{\textrm{SNstep}@\omega_{7p,4s}}^{(2)}(\omega_{4s,3s} - \delta)$ is also a
real quantity that, for a given spectrum, its positive or negative sign depends on $\delta$.
This is seen in Fig.~\ref{fig5}, where the difference from the TL case is clearly observable.
Since the TL pulse induces fully constructive interferences among all the two-photon pathways
that contribute to $A^{(2)}(\Omega)$ for any given $\Omega$, with a given spectrum, the TL value of
$A^{(2)}_{\textrm{TL}}(\omega_{4s,3s}-\delta)$ is actually the maximal one for any given $\delta$.
As such, it serves as an upper limit to the value of
$\left|A_{\textrm{SNstep}@\omega_{7p,4s}}^{(2)}(\omega_{4s,3s} - \delta)\right|$.

As can be seen from Fig.~\ref{fig5}, the values of $A_{\textrm{SNstep}@\omega_{7p,4s}}^{(2)}(\omega_{4s,3s} - \delta)$
at $\delta=0$ and at $\delta\approx0$ strongly depend on the spectral bandwidth:
with 5-nm and 9-nm bandwidth the values for $\delta=0$ and $\delta\approx0$ are all positive [panels (a) and (b)],
with 15-nm bandwidth the value for $\delta=0$ is almost zero while the values for small
$\pm\left|\delta\right|$ are of different signs [panel (c)],
and with 25-nm and 40-nm bandwidth the values for $\delta=0$ and $\delta\approx0$ are all negative [panels (d) and (e)].
On the other hand, with any spectral bandwidth, the field
$\widetilde{E}_{\textrm{SNstep}@\omega_{7p,4s}}(\omega_{7p,4s}+\delta)$
has different sign for +$\left|\delta\right|$ and $-$$\left|\delta\right|$,
which is in full correlation with the corresponding sign of $1/\delta$.
Thus, the relative sign between the amplitudes
$\frac{1}{\delta} A^{(2+1)}_{\textrm{SNstep}@\omega_{7p,4s}}(\delta)$
contributed by $\pm\left|\delta\right|$
to the $\wp$-integration of $A_{7p,\textrm{SNstep}@\omega_{7p,4s}}^{(2+1)near-res}$
is determined only by $A^{(2)}_{\textrm{SNstep}@\omega_{7p,4s}}(\omega_{4s,3s}-\delta)$.
As such, it depends here on the spectral bandwidth.
In Fig.~\ref{fig5}, the $A^{(2)}_{\textrm{SNstep}@\omega_{7p,4s}}(\omega_{4s,3s}-\delta)$ values
for $\pm\left|\delta\right|$ are of the same sign,
i.e., they interfere constructively one with the other,
in all the bandwidth cases except for the 15-nm bandwidth,
where they are of different sign and thus interfere destructively one with the other.
The constructive nature of these interferences generally leads to the dominance of the corresponding near-resonant
component $A_{7p,\textrm{SNstep}@\omega_{7p,4s}}^{(2+1)near-res}$ over the on-resonant
component $A_{7p,\textrm{SNstep}@\omega_{7p,4s}}^{(2+1)on-res}$ for all bandwidth values, except for
values around 15~nm. There, both the near-resonant and on-resonant components are very small and
have a value close to zero.
This is indeed the behavior of the near-resonant vs.~on-resonant components shown in Fig.~\ref{fig4}(a).

For analyzing the bandwidth dependence of the absorption relative to the TL pulse,
i.e., the graph of $\widetilde{P}_{7p,\textrm{SNstep}@\omega_{7p,4s}}^{(2+1)}$ vs.~$\Delta\omega$ shown in Fig.~\ref{fig4}(a) (solid line),
one also needs to account for the case of the TL pulse analyzed in the previous section.
At a very narrow bandwidth the phase shaping is expected to have
very small or no effect at all on the absorption relative to the
(unshaped) TL pulse, as is indeed reflected in the value
$\widetilde{P}_{7p,\textrm{SNstep}@\omega_{7p,4s}}^{(2+1)} = 1$ for
2-nm bandwidth.
Then, as the bandwidth increases the pulse shaping starts to have an effect.
For the 5-nm bandwidth,
where there is a local peak of $\widetilde{P}_{7p,\textrm{SNstep}@\omega_{7p,4s}}^{(2+1)} \sim 2.5$,
one obtains that:
(i) $\left|A_{7p,{\textrm{SNstep}@\omega_{7p,4s}}}^{(2+1)on-res}\right| \approx \left|A_{7p,\textrm{TL}}^{(2+1)on-res}\right|$
since $A^{(2)}_{\textrm{SNstep}@\omega_{7p,4s}}(\omega_{4s,3s}-\delta) \approx A^{(2)}_{\textrm{TL}}(\omega_{4s,3s}-\delta)$
[see Fig.~\ref{fig5}(a)], and
(ii) $\left|A_{7p,\textrm{SNstep}@\omega_{7p,4s}}^{(2+1)near-res}\right| > \left|A_{7p,\textrm{TL}}^{(2+1)near-res}\right|$
due to the different nature of the interferences between positively- and negatively-detuned excitation pathways for the two pulses.
Hence,
$\widetilde{P}_{7p,\textrm{SNstep}@\omega_{7p,4s}}^{(2+1)} = \frac
{\left|A_{7p,\textrm{SNstep}@\omega_{7p,4s}}^{(2+1)on-res}\right|^{2} +
 \left|A_{7p,\textrm{SNstep}@\omega_{7p,4s}}^{(2+1)near-res}\right|^{2}}
{\left|A_{7p,\textrm{TL}}^{(2+1)on-res}\right|^{2} + \left|A_{7p,\textrm{TL}}^{(2+1)near-res}\right|^{2}}  > 1 \; $
for 5-nm bandwidth.
Then, as the bandwidth increases from 5~nm to 9~nm,
the value of $\widetilde{P}_{7p,\textrm{SNstep}@\omega_{7p,4s}}^{(2+1)}$ decrease to a value around 1.
This decrease results mainly from the stronger increase of
$A^{(2)}_{\textrm{TL}}(\omega_{4s,3s})$ relative to $A^{(2)}_{\textrm{SNstep}@\omega_{7p,4s}}(\omega_{4s,3s}-\delta)$
for $\delta=0$ and $\delta\approx 0$ [see Fig.~\ref{fig5}(b)].
%
Then, as the bandwidth further increases to around 15 nm, the value of
$\widetilde{P}_{7p,\textrm{SNstep}@\omega_{7p,4s}}^{(2+1)}$ approach zero
since $\left|A_{7p,{\textrm{SNstep}@\omega_{7p,4s}}}^{(2+1)on-res}\right| \approx 0$ and
$\left|A_{7p,{\textrm{SNstep}@\omega_{7p,4s}}}^{(2+1)near-res}\right| \approx 0$,
resulting from
$A_{\textrm{SNstep}@\omega_{7p,4s}}^{(2)}(\omega_{4s,3s}) \approx 0$ and
$A_{\textrm{SNstep}@\omega_{7p,4s}}^{(2)}(\omega_{4s,3s}-\delta)\approx 0$ for $\delta\approx0$
[see above and Fig.~\ref{fig5}(c)].
Actually, with 14.6-nm bandwidth the on-resonant component is completely zeroing out and
to a very good approximation the corresponding shaped pulse is a three-photon dark pulse.

Further increase of the bandwidth beyond 15~nm, leads to a continuous increase in the
value of $\widetilde{P}_{7p,\textrm{SNstep}@\omega_{7p,4s}}^{(2+1)}$.
For example, at 40-nm bandwidth it reaches a value of 4.1.
Following the above analysis and as seen in Fig.~\ref{fig3}(d1)-(e1)
and Fig.~\ref{fig4}(a), for bandwidth values beyond 15~nm, the value
of $\widetilde{P}_{7p,\textrm{SNstep}@\omega_{7p,4s}}^{(2+1)}$ is
actually determined only by the dominant near-resonant
shaped-pulse term
$\left|A_{7p,{\textrm{SNstep}@\omega_{7p,4s}}}^{(2+1)near-res}\right|$
and the on-resonant TL term
$\left|A_{7p,{\textrm{TL}}}^{(2+1)on-res}\right|$, i.e.,
$\widetilde{P}_{7p,\textrm{SNstep}@\omega_{7p,4s}}^{(2+1)} \approx
\left|A_{7p,{\textrm{SNstep}@\omega_{7p,4s}}}^{(2+1)near-res}\right|^{2}
/ \left|A_{7p,{\textrm{TL}}}^{(2+1)on-res}\right|^{2}$.
For large bandwidth $\Delta\omega$, $\widetilde{E}(\omega_{7p,4s}) \approx \widetilde{E}(\omega_{0}) \equiv 1$
and $\widetilde{E}(\omega_{7p,4s}-\delta) \approx \widetilde{E}(\omega_{0}) \equiv 1$ for small $\delta$.
Hence:
(i) For the TL pulse:
since $A^{(2)}_{\textrm{TL}}(\omega_{4s,3s}) \propto \Delta\omega$,
based on Eq.~(\ref{eq_Af_on_res}), $\left|A_{7p,{\textrm{TL}}}^{(2+1)on-res}\right| = k_{1} \Delta\omega$,
where $k_{1}$ is a proportionality constant for the TL pulse;
(ii) For the shaped pulse of $\omega^{\textrm{(SNstep)}}_{\textrm{step}} = \omega_{7p,4s}$:
$A_{7p,{\textrm{SNstep}@\omega_{7p,4s}}}^{(2+1)near-res} =
- \wp\int_{-\infty}^{\infty} \frac{1}{\delta} A^{(2)}(\omega_{4s,3s}-\delta) d\delta =
k_{2a}\Delta\omega + k_{2b} \Delta\omega \ln{\Delta\omega}$, where $k_{2a}$ and $k_{2b}$ are constants
for the specific single-step pulse shape.
As a result, for large enough bandwidth values, an increase in the bandwidth $\Delta\omega$
leads to a continuous increase in the value
of $\widetilde{P}_{7p,\textrm{SNstep}@\omega_{7p,4s}}^{(2+1)}$ as $K_{2a} + K_{2b}\ln{\Delta\omega}$.
This is indeed seen in the graph shown in Fig.~\ref{fig4}(a) (solid line).

\subsection{Shaped pulse with double-step phase pattern of $\omega^{\textrm{(DBstep)}}_{\textrm{left-step}} = \omega_{7p,4s}$}

Last we consider the case of the shaped pulse with the phase pattern of a double $\pi$ step with
$\omega^{\textrm{(DBstep)}}_{\textrm{left-step}} = \omega_{7p,4s}$ (solid thin black lines in Fig.~\ref{fig5}).
With such a phase pattern, as with any of the double-step phase patterns,
the phase associated with all the two-photon pathways of total two-photon transition frequency $\omega_{4s,3s}$
is zero, since $\Phi(\omega) + \Phi(\omega_{4s,3s}-\omega) = 0$ for any $\omega$ (see Fig.~\ref{fig1}).
So, fully constructive interferences are induced between all these two-photon pathways, as is the
case with the transform-limited pulse.
Hence, as seen in Fig.~\ref{fig5}, with any spectral bandwidth,
$A^{(2)}_{\textrm{DBstep}@\omega_{7p,4s}}(\omega_{4s,3s}) = A^{(2)}_{\textrm{TL}}(\omega_{4s,3s})$,
which, as explained above, is the maximal possible value (here, at $\delta=0$) for a given spectrum.
This leads to a maximal on-resonant amplitude that is of equal magnitude to the TL case, i.e.,
$\left|A_{7p,\textrm{DBstep}@\omega_{7p,4s}}^{(2+1)on-res}\right| = \left|A_{7p,\textrm{TL}}^{(2+1)on-res}\right|$,
for any bandwidth $\Delta\omega$.

With the double-step phase pattern of $\omega^{\textrm{(DBstep)}}_{\textrm{left-step}} = \omega_{7p,4s}$,
the field $\widetilde{E}_{\textrm{DBstep}@\omega_{7p,4s}}(\omega_{7p,4s}+\delta)$
has different sign for $\pm\left|\delta\right|$ when $\left|\delta\right|$ is small,
which is in full correlation with the sign of $1/\delta$ for small $\pm\left|\delta\right|$.
This is similar to the case of the single-step phase pattern of $\omega^{\textrm{(SNstep)}}_{\textrm{step}} = \omega_{7p,4s}$ discussed above.
Thus, also here,
the relative sign between the amplitudes
$\frac{1}{\delta} A^{(2+1)}_{\textrm{DBstep}@\omega_{7p,4s}}(\delta)$
contributed to the $\wp$-integration of $A_{7p,\textrm{DBstep}@\omega_{7p,4s}}^{(2+1)near-res}$ from $\pm\left|\delta\right|$
is determined only by $A^{(2)}_{\textrm{DBstep}@\omega_{7p,4s}}(\omega_{4s,3s}-\delta)$.

The fact that $A^{(2)}_{\textrm{DBstep}@\omega_{7p,4s}}(\omega_{4s,3s})$ (i.e., for $\delta=0$)
has the maximal possible value also implies that
for small $\delta$ values $A^{(2)}_{\textrm{DBstep}@\omega_{7p,4s}}(\omega_{4s,3s}-\delta)$
is also very close to the maximal possible TL value, i.e.,
$A^{(2)}_{\textrm{DBstep}@\omega_{7p,4s}}(\omega_{4s,3s}-\delta) \approx A^{(2)}_{\textrm{TL}}(\omega_{4s,3s}-\delta)$,
with a corresponding value that is a real positive one.
As seen in Fig.~\ref{fig5}, for any bandwidth,
$A^{(2)}_{\textrm{DBstep}@\omega_{7p,4s}}(\omega_{4s,3s}-\delta)$ is of the same sign for
small $\pm\left|\delta\right|$,
i.e., the interferences between negatively- and positively-detuned near-resonant pathways are constructive ones.
%
Similar to the single-step phase pattern discussed above,
the constructive nature of these interferences leads to the dominance of the corresponding near-resonant
component $A_{7p,\textrm{DBstep}@\omega_{7p,4s}}^{(2+1)near-res}$ over the on-resonant
component $A_{7p,\textrm{DBstep}@\omega_{7p,4s}}^{(2+1)on-res}$ for all bandwidth values.
This is indeed the behavior of the near-resonant vs.~on-resonant components shown in Fig.~\ref{fig4}(b).
The fact that
$A^{(2)}_{\textrm{DBstep}@\omega_{7p,4s}}(\omega_{4s,3s}-\delta) \approx A^{(2)}_{\textrm{TL}}(\omega_{4s,3s}-\delta)$,
while the nature of the near-resonant interferences is constructive to the pulse with the double-step
phase pattern and destructive for the TL pulse, imply that
$\left|A_{7p,\textrm{DBstep}@\omega_{7p,4s}}^{(2+1)near-res}\right|^{2} \gg \left|A_{7p,\textrm{TL}}^{(2+1)near-res}\right|^{2}$.

The overall bandwidth dependence of $\widetilde{P}_{7p,\textrm{DBstep}@\omega_{7p,4s}}^{(2+1)}$ shown in Fig.~\ref{fig4}(b) (solid line),
can then be understood as follows:
Since
$\left|A_{7p,\textrm{DBstep}@\omega_{7p,4s}}^{(2+1)on-res}\right|^{2} = \left|A_{7p,\textrm{TL}}^{(2+1)on-res}\right|^{2}$
and
$\left|A_{7p,\textrm{DBstep}@\omega_{7p,4s}}^{(2+1)near-res}\right|^{2} > \left|A_{7p,\textrm{TL}}^{(2+1)near-res}\right|^{2}$
one obtains that $\widetilde{P}_{7p,\textrm{DBstep}@\omega_{7p,4s}}^{(2+1)} > 1$ for any bandwidth value.
This is indeed what observed in Fig.~\ref{fig4}(b),
i.e., the corresponding three-photon absorption is always above the TL absorption.

The asymptotic behavior of the graph of
$\widetilde{P}_{7p,\textrm{DBstep}@\omega_{7p,4s}}^{(2+1)}$-vs.-$\Delta\omega$
is actually very similar to the single-step phase pattern case analyzed above, just with
higher values due to the fact that $A^{(2)}_{\textrm{DBstep}@\omega_{7p,4s}}(\omega_{4s,3s})$ is kept on its maximal value for
any bandwidth.
Also here, for large bandwidth $\Delta\omega$,
following the above analysis for the present double-phase-step pulse and for the TL pulse,
$\widetilde{P}_{7p,\textrm{DBstep}@\omega_{7p,4s}}^{(2+1)} \approx \left|A_{7p,{\textrm{DBstep}@\omega_{7p,4s}}}^{(2+1)near-res}\right|^{2} /
\left|A_{7p,{\textrm{TL}}}^{(2+1)on-res}\right|^{2}$.
Also, $\widetilde{E}(\omega_{7p,4s}) \approx \widetilde{E}(\omega_{0}) \equiv 1$
and $\widetilde{E}(\omega_{7p,4s}-\delta) \approx \widetilde{E}(\omega_{0}) \equiv 1$ for small $\delta$.
So:
(i) For the TL pulse:
$A^{(2)}_{\textrm{TL}}(\omega_{4s,3s}) \propto \Delta\omega$ and thus
$\left|A_{7p,{\textrm{TL}}}^{(2+1)on-res}\right| = k_{1} \Delta\omega$, where $k_{1}$ is
the TL proportionality constant.
(ii) For the shaped pulse of $\omega^{\textrm{(DBstep)}}_{\textrm{left-step}} = \omega_{7p,4s}$:
$A_{7p,{\textrm{DBstep}@\omega_{7p,4s}}}^{(2+1)near-res} =
- \wp\int_{-\infty}^{\infty} \frac{1}{\delta} A^{(2)}(\omega_{4s,3s}-\delta) d\delta =
k_{3a}\Delta\omega + k_{3b} \Delta\omega \ln{\Delta\omega}$, where $k_{3a}$ and $k_{3b}$ are
constants for the specific double-phase-step pulse shape.
So, $\widetilde{P}_{7p,\textrm{DBstep}@\omega_{7p,4s}}^{(2+1)}$ behaves asymptotically
as $K_{3a} + K_{3b}\ln{\Delta\omega}$.
The constants $K_{3a}$ and $K_{3b}$ are, respectively, larger than the $K_{2a}$ and $K_{2b}$ constants
of the single-step phase pattern of $\omega^{\textrm{(SNstep)}}_{\textrm{step}} = \omega_{7p,4s}$,
due to the larger near-resonant component induced by the double-step phase pattern.
Hence, for any given bandwidth,
the resonance-mediated (2+1) three-photon absorption induced by the
shaped pulse with double-step phase pattern of $\omega^{\textrm{(DBstep)}}_{\textrm{left-step}} = \omega_{7p,4s}$
is higher than the absorption induced by the
shaped pulse with single-step phase pattern of $\omega^{\textrm{(SNstep)}}_{\textrm{step}} = \omega_{7p,4s}$.

\section{\label{sec:conclusions}Conclusions}

In conclusion, we have studied in detail 
femtosecond phase control of atomic resonance-mediated (2+1)
three-photon absorption, focusing on its 
dependence on the spectral bandwidth of the excitation pulse.
The total absorption amplitude has contributions from a group of interfering three-photon pathways that are on resonant
with the intermediate state and a group of interfering three-photon pathways that are near resonant with it.
For a given pulse shape, the ratio between the amplitude contributions from these two groups
depends on the spectral excitation bandwidth.
The nature of the three-photon absorption that is induced by a given
shaped pulse is analyzed in a systematic and physically-intuitive
way by decomposing it into a two-photon absorption, which is induced
by all the possible two-photon pathways leading to various detunings
from the intermediate state, followed by the absorption of a proper
third photon.
Prominent among the results is our finding that with simple proper pulse shaping
an increase in the excitation bandwidth leads to a corresponding increase in the enhancement of the three-photon
absorption over the absorption induced by the (unshaped) transform-limited pulse.
For example, here, 40-nm bandwidth leads to an order-of-magnitude enhancement over the transform-limited absorption.
The present work serves as a basis for future extensions to
molecular systems and we expect it to be significant for nonlinear
spectroscopy and microscopy.


\section*{ACKNOWLEDGMENTS}
This research was supported by The Israel Science Foundation (grant No. 127/02),
by The James Franck Program in Laser Matter Interaction,
and by The Technion's Fund for The Promotion of Research.

\newpage 


\newpage


\end{document}